\documentclass[preprint,showpacs,preprintnumbers,amsmath,amssymb]{revtex4}
\usepackage{graphicx}
\usepackage{dcolumn}
\usepackage{bm}
\usepackage{amssymb}
\usepackage{amsfonts}
\usepackage{amsmath}
\usepackage{textcomp}
\begin{document}

\title{Conditions for the generation of spin currents in spin-Hall devices}
\author{J.-E. Wegrowe$^1$} \email{jean-eric.wegrowe@polytechnique.edu} \author{R. V. Benda$^{1,2}$} \author{J. M. Rub\`i$^{2}$} 
\affiliation{$^1$ Ecole Polytechnique, LSI, CNRS and CEA/DSM/IRAMIS, Palaiseau F-91128, France} 
\affiliation{$^2$ Departament de F\'isica Fonamental, Fascultat de F\'isica, Universitat de Barcelona, 08028, Spain}

\date{\today}

\begin{abstract}
We investigate the out-of-equilibrium stationary states of spin-Hall devices on the basis of the least dissipation principle. We show that, for a bulk paramagnet with spin-orbit interaction, in the case of the Hall bar geometry the principle of minimum dissipated power prevents the generation of transverse spin and charge currents while in the case of the Corbino disk geometry, transverse currents can be produced. More generally, we show that electric charge accumulation prevents the stationary spin to charge current conversion.
\end{abstract}
\maketitle 

\vspace{2.5cm}

\section{Introduction}

The interest for low power consumption electronic devices is one of the main motivations for the research in spintronics. In this context, considerable efforts have been devoted to the study of the spin Hall effect (SHE) and the inverse spin Hall effect (ISHE) \cite{Dyakonov1,Hirsch,Zhang, Zutic,Maekawa,Dyakonov2,Hoffmann,Zhang14,Sinova,Saslow}. In the Dyakonov-Perel (DP) model \cite{Dyakonov1,Dyakonov2}, SHE is a generalization of the usual Hall effect in which the spin-orbit interaction plays the role of an internal magnetic field that depends on the electronic spin states. In particular, SHE describes the spin-accumulation produced in the direction perpendicular to an injected current. The inverse effect ISHE refers to the experimental configuration in which a spin current - or a non-uniform spin-accumulation- generates a transverse electric field. These effects open the way to the implementation of new devices, that allow the conversion of spin current into charge current \cite{Patent}. 
Establishing the conditions under which this conversion, that takes place in a stationary state, obeys the thermodynamic laws is a problem of crucial importance currently.

Both SHE and ISHE can be analyzed by the spin-Hall transport equations, first proposed by Dyakonov and Perel \cite{Dyakonov1} in 1971 to describe paramagnetic conductors with spin-orbit interaction. A similar effect is well-known for bulk ferromagnetic conductors in terms of an anomalous Hall effect (AHE) \cite{Nagaosa}. After the discovery of giant magnetoresistance at the turn of the millennium, SHE was also considered at an interface in the context of spin-injection \cite{Hirsch,Zhang}. In parallel, the concept of pure spin current was proposed \cite{Jedema,Georges} in order to describe spin-accumulation occurring in a part of the device that does not drive electric current. The presence of a spin-current was found in a great diversity of systems and situations \cite{Maekawa,Hoffmann,Sinova,Otani,Sonin}. The conditions for its existence is an open problem crucial in the energetics of the system since that current, in contrast to spin-accumulation, entails heat dissipation.

The goal of this work is to calculate the minimum power dissipation in different kinds of spin-Hall devices, in order to deduce the conditions for the existence of pure spin-current. The study is first restricted to the case of the DP model described in the framework of the two spin channel approach. We will show that 
in the case of the Hall bar geometry the principle of minimum dissipated power prevents the generation of transverse spin and charge currents, while in the case of the Corbino disk geometry, a transverse pure spin-current is possible. 

In the framework of the two channel model, the system is defined by an ensemble of two populations of electric carriers: those with spins $\uparrow$ and $\downarrow$ ; the spin-polarization axis is defined by a unit vector $\vec p$ perpendicular to the current.  We assume that the temperature is constant everywhere. 
The transport equations are then described by the Ohm's law applied to each spin-channel:
\begin{equation}
\vec J_{\updownarrow} = - \hat \sigma_{\updownarrow} \vec \nabla \mu_{\updownarrow}, 
\label{Ohm1}
\end{equation}
where $\hat \sigma_{\updownarrow}$ is the conductivity tensor and $\mu_{\updownarrow}$ is the electrochemical potential. We assume for convenience that the system is a two-dimensional thin layer so that $\hat \sigma_{\updownarrow}$ is a 4x4 matrix, which is isotropic in the absence of spin-orbit coupling (see reference \cite{AThT} for the case of a ferromagnetic material). In a cartesian coordinate system $\{\vec e_x, \vec e_y \}$ we have $\sigma_{xx \updownarrow}=\sigma_{yy \updownarrow} =\sigma_\updownarrow$.

The Dyakonov-Perel (DP) equations are obtained by the application of the relevant symmetries to the spin-dependent conductivity tensor $\hat \sigma_{\updownarrow}$. Since, in the DP model, the effect of the spin-orbit coupling is equivalent to that of a local magnetic field perpendicular to the current, the cross-coefficients obey the Onsager reciprocity relations \cite{Onsager1,Onsager2} $\sigma_{xy \updownarrow} = - \sigma_{yx\updownarrow} = \sigma_{so \updownarrow}$ (where $\sigma_{so \updownarrow}$ is the Hall conductivity due to this local spin-orbit field). 
Furthermore, if the local spin-orbit field due to the spin-polarization $\uparrow$ is along the direction $\vec e_z$ the spin-orbit field due to the spin-polarization $\downarrow$ is along the direction $- \vec e_z$, so that the Onsager-Casimir relations imposes $\sigma_{so \uparrow} = - \sigma_{so \downarrow}$. In the Cartesian reference frame, the conductivity tensor reads : 
\begin{equation}
\hat \sigma_{\updownarrow} =
\left( \begin{array}{cccc}
                       \sigma_{\uparrow} & \sigma_{so} & 0 & 0  \\
                      - \sigma_{so}  &  \sigma_{\uparrow} & 0 & 0 \\
                      0 & 0 &  \sigma_{\downarrow} & -\sigma_{so} \\
                      0 & 0 & \sigma_{so}  &  \sigma_{\downarrow} \\
		\end{array} \right).                
  \label{Conductivity}
\end{equation}
The resistivity tensor $\hat \rho = \hat \sigma^{-1}$ is hence defined by the two coefficients $\rho_{\updownarrow} = \sigma_{\updownarrow}/(\sigma_{\updownarrow}^2 + \sigma_{so}^2)$ and  $\rho_{so} = \sigma_{so}/(\sigma_{\updownarrow}^2 + \sigma_{so}^2)$.
In order to take into account the diffusion of electric charges, the density of spin-dependent carriers $n_{\updownarrow}$ are not necessarily constant throughout the material, and the corresponding diffusion terms should be taken into account in the electrochemical potential \cite{Rubi}:
\begin{equation}
\mu_{\updownarrow} = \frac{kT_F}{q} \ln(n_{\updownarrow})  + V + \mu^{ch}_{\updownarrow},
\label{ChemPot} 
\end{equation}
where $T_F$ is either the Fermi temperature in the case of a metal or the temperature of the thermostat in the case of a semi-conductor (non-degenerated). The Boltzmann constant is $k$, $V$ the electrostatic potential, $\mu^{ch}_{\updownarrow}$ is the spin-dependent chemical potential, and $q$ is the charge of the carriers. The gradient of each term of the electrochemical potential is a force : the thermodynamic force proper of the diffusion process $ \vec \nabla n_{\updownarrow}$, the electric field $\vec {\mathcal E}_{\updownarrow} \equiv - \vec \nabla (V + \mu^{ch}_{\updownarrow}$) and the spin-flip force $\Delta  \mu^{ch}$ \cite{MagDiff,Entropy}. Introducing Eq.(\ref{ChemPot}) and Eq.(\ref{Conductivity}) in Eq.(\ref{Ohm1}) we have:
\begin{equation}
 \begin{array}{c}
\vec J_{\updownarrow} =  -  (\sigma_{\updownarrow} \, \vec {\mathcal E}_{\updownarrow} + D_{\updownarrow} \vec \nabla n_{\updownarrow}) \pm  \, \vec p \times (\sigma_{so}\, \vec {\mathcal E}_{\updownarrow} + D_{so \updownarrow} \vec \nabla n_{\updownarrow})
\end{array}
\label{Current2}
\end{equation}
where the signs $+$ and $-$ correspond to channels $\uparrow$ and $\downarrow$ respectively. We have introduced the longitudinal and transverse spin-dependent diffusion coefficients $D_{\updownarrow} = kT_F \sigma_{\updownarrow}/(q n_{\updownarrow})$ and  $D_{so \updownarrow} = kT_F \sigma_{so}/(qn_{\updownarrow})$. The current of electric charges is defined by the sum $\vec J_c = \vec J_{\uparrow} + \vec J_{\downarrow}$, and the spin-current is defined by the difference $\vec J_s = \vec J_{\uparrow} - \vec J_{\downarrow}$. Equations (\ref{Current2}) are equivalent or generalize those derived in references \cite{Zhang,Zutic}. 
In a first step, we restrict the analysis to the DP model without spin-flip scattering. In this case $\mu^{ch}_{\uparrow}$ and $\mu^{ch}_{\downarrow}$ are constant in space and the electric field is spin-independent: $ \vec{\mathcal E}_{\updownarrow} = \vec{\mathcal E} = - \vec \nabla V$. If in addition we assume that there is no conductivity asymmetry ($\sigma_{\uparrow} = \sigma_{\downarrow} = \sigma$), the equations (\ref{Current2}) are then equivalent to those of Dyakonov and Perel \cite{Dyakonov1,Dyakonov2,Saslow} (as shown in reference \cite{SPIE14}). 
\begin{figure} [h!]
   \begin{center} 
   \begin{tabular}{c}
  \includegraphics[height=5cm]{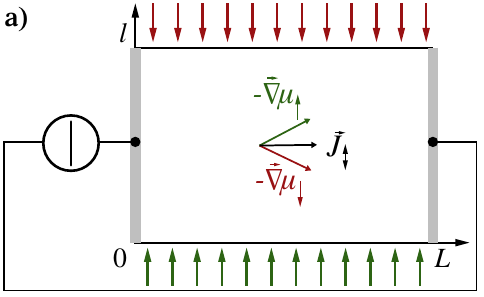}
  \includegraphics[height=5cm]{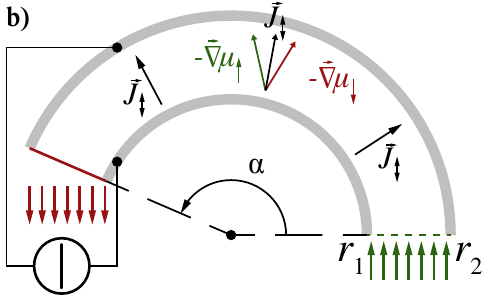}
   \end{tabular}
   \end{center}
   \caption[Fig1]
{ \label{fig:Fig1} : Schematic views of two equivalent spin-Hall devices: (a) Hall-bar and (b) Corbino sector. The spin-Hall angles $\theta_{SH \updownarrow}$ are defined as the angles between $- \vec \nabla \mu_{\updownarrow}$ and the injected current $\vec J_{\updownarrow}$. }
   \end{figure}
   We now apply the Kirchhoff-Helmoltz principle which states that the current distributes itself so as to minimize heat dissipation, for a given bias voltage. 
In other terms, the stationary states are those states at which dissipation is a minimum compatible with the ensemble of constraints applied to the system \cite{Onsager3,Jaynes}.
In the case of Hall devices, the most important constraint that imposes non-equilibrium is the injection of the electric current (or the application of a voltage difference) along one direction of the Hall device (along the $ \vec e_x$ direction in the case of the Hall bar). Fields and charge densities are governed by Maxwell equations which can be viewed as constraints to the values of these quantities. 

The power dissipated by the system is $P_J = \vec J_{\uparrow} .\vec \nabla \mu_{\uparrow} + \vec J_{\downarrow} .\vec \nabla \mu_{\downarrow}$. In view of the linear relation (\ref{Ohm1}) it is a function of the partial derivatives of the electrochemical potentials $\partial_j \mu_{\updownarrow}$ (where $j=\{x,y\}$) (that contain drift and diffusion terms) and a function of the charge accumulation $\partial_j n_{\updownarrow}$ (electrostatic forces). The power can equivalently be expressed as a function of the currents $\vec J_{\updownarrow}$ or the thermodynamic forces, through Eqs.(\ref{Ohm1}).  The Coulomb equation
$div( \vec{\mathcal E} )= q n/\epsilon$ ($\epsilon$ is the permittivity and $n = n_{\uparrow}+ n_{\downarrow}$ the total electric charge at each point) and the expression $2 \vec {\mathcal E} = - \vec \nabla \mu_{\uparrow} + c_{\uparrow} \vec \nabla n_{\uparrow} - \vec \nabla \mu_{\downarrow}  +c_{\downarrow} \vec \nabla n_{\downarrow} $ lead to the Poisson law $\nabla^2( \mu_{\uparrow} +  \mu_{\uparrow} ) - c_{\uparrow} \nabla^2 n_{\uparrow} - c_{\downarrow} \nabla^2 n_{\downarrow} + 2q n/\epsilon = 0  $, where $c_{\updownarrow} = kT_F/(q n_{\updownarrow})$. We do not impose any boundary conditions on the currents at the edges of the Hall bar along $\vec e_y$ (e.g. a wire can connect the two opposite edges) \cite{Rque}. In order to find the minimum of the Joule power $P_J$ compatible with the constraints, we introduce the corresponding Lagrange multiplayers $\lambda(x,y), \beta(y)$ such that the function to minimize reads:
\begin{eqnarray}
&& \mathcal F [\vec \nabla \mu_{\updownarrow}, n_{\updownarrow}, \lambda, \beta] = \int_0^L \int_0^L [ \nonumber \\
&&  \sigma_{\uparrow} \left(\partial_x \mu_{\uparrow} \right)^2 + \sigma_{\uparrow} \left(\partial_y \mu_{\uparrow} \right)^2 +\sigma_{\downarrow} \left(\partial_x \mu_{\downarrow} \right)^2 + \sigma_{\downarrow} \left(\partial_y \mu_{\downarrow}\right)^2  
\nonumber \\
&& - \lambda(x,y) \left(\nabla^2 \left( \mu_{\uparrow}+\mu_{\downarrow} \right) -c_{\uparrow} \nabla^2 n_{\uparrow} -c_{\downarrow} \nabla^2 n_{\downarrow}+2q \frac{n}{\epsilon} \right) \nonumber \\
&& - \beta(y) ( -\sigma_{\uparrow} \partial_x \mu_{\uparrow} -\sigma_{so} \partial_y \mu_{\uparrow} \nonumber \\
&& + \epsilon_i \left( - \sigma_{\downarrow} \partial_x \mu_{\downarrow} +\sigma_{so} \partial_y \mu_{\downarrow} \right)-J_{i x}^{\circ}  )
\, \,  ]  \, dx dy,
\label{F_Min}
\end{eqnarray}
where $J_{i x}^{\circ} = \vec J_i^{\circ} . \vec e_x$ is the constant charge or spin current density injected in the device along the $x$ direction,
with the different options: $i = \{c, s\}$ (with $i= c$ associated to a charge current  and $\epsilon_c=1$, $i= s$ associated to a spin current and $\epsilon_s=-1$), and $\vec \nabla = (\partial_x, \partial_y)$. In the following equations, $\pm$ or $\mp$ account for the differences between the two spin channels $\uparrow$ and $\downarrow$. Minimization of $\mathcal F$ in the case of an imposed charge current ($i=c$) leads to the following conditions: 
\begin{eqnarray}
\nabla_{\partial_x \mu_{\updownarrow}} \mathcal F= 0 \Rightarrow \partial_x \mu_{\updownarrow}=-\frac{1}{2} \beta(y)-\frac{1}{2\sigma_{\updownarrow}} \partial_x \lambda \nonumber \\
\nabla_{\partial_y \mu_{\updownarrow}} \mathcal F = 0 \Rightarrow \partial_y \mu_{\updownarrow}= \mp \frac{1}{2} \frac{\sigma_{so}}{\sigma_{\updownarrow}} \beta(y)-\frac{1}{2\sigma_{\updownarrow}} \partial_y \lambda 
\label{Min_eqs}
\end{eqnarray}
and:
\begin{eqnarray}
 \nabla_{n_{\updownarrow}} \mathcal F =c_{\updownarrow} \nabla^2 \lambda -2\frac{q}{\epsilon}  \lambda=0 
 \label{eq_bulk_total_current_cst6}
\end{eqnarray}
The solution for the chemical potential in the case $i=c$, derived from (\ref{Min_eqs}) and (\ref{eq_bulk_total_current_cst6}) is :(see \cite{SuppMat})
\begin{equation}
\lambda_D^2 \nabla^2 \mu_{\updownarrow} = 2 \mu_{\updownarrow}(x,y) + \beta  (x \pm \frac{\sigma_{so}}{\sigma_{\updownarrow}}y ) \, +C_{\updownarrow}
\label{Result0}
\end{equation}
where $\lambda_D = \sqrt{\epsilon kT_F/(q^2n)}$ is the Debye-Fermi length, $C_{\updownarrow}$ is a constant and $\beta(y) = \beta$ is the Lagrange multiplayer introduced above. Note that Eq.(\ref{Result0}) can be interpreted as a screening equation for the electric charges, with the term $\beta (x \pm \frac{\sigma_{so}}{\sigma_{\updownarrow}}y ) $ as a source term. We assume $\lambda_D \ll l$, where $l$ is the typical length along the direction $\vec e_y$. The corresponding solution of Eq.(\ref{Result0}) \textit{in the bulk} is then : 
\begin{equation}
\mu_{\updownarrow}(x,y) =-(x \pm \frac{\sigma_{so}}{\sigma_{\updownarrow}}y) \, \tilde \rho J_{cx }^{o}- \frac{C_{\updownarrow}}{2}
\label{Result}
\end{equation}
with $\tilde \rho = \frac{ \sigma_{\uparrow}\sigma_{\downarrow}}{(\sigma_{\uparrow}+\sigma_{\downarrow}) \left ( \sigma_{\uparrow}\sigma_{\downarrow}+ \sigma_{so}^2 \right ) } = \left(\frac{1}{\rho_{\uparrow}}+\frac{1}{\rho_{\uparrow}} \right)^{-1}$ the mean resistivity and according to Eq.(\ref{ChemPot}), the constant is $\mu^{ch}_{\updownarrow}$. 
In the case $i=s$ of an imposed spin current ($\epsilon_i=-1$ in Eq. (\ref{F_Min})), we find :
\begin{equation}
\lambda_D^2 \nabla^2 \mu_{\updownarrow} = 2 \mu_{\updownarrow}(x,y) \pm \beta  (x \pm \frac{\sigma_{so}}{\sigma_{\updownarrow}}y ) \, +\tilde{C}_{\updownarrow},
\label{Result0_i=s}
\end{equation}
(where $\tilde{C}_{\updownarrow}$ is a constant) which gives in the bulk ($l >> \lambda_D$) :
\begin{equation}
\mu_{\updownarrow}(x,y) = \mp (x \pm \frac{\sigma_{so}}{\sigma_{\updownarrow}}y) \, \tilde \rho J_{sx }^{o}-\frac{\tilde{C}_{\updownarrow}}{2}.
\label{Result_i=s}
\end{equation}
Inserting Eq.(\ref{Result}) or (\ref{Result_i=s}) into Eq.(\ref{Ohm1}) we obtain: 
\begin{equation}
\vec J_{\updownarrow}. \vec e_y = J_{y \updownarrow} = 0
\label{Zero}
\end{equation}
or $J_{c y} =J_{s y} = 0$. For $i=c$ : $J_{x \updownarrow} = \, (\tilde \rho/\rho_{\updownarrow}) J_{c x}^0$ and for $i=s$ : $J_{x \updownarrow} = \pm \, (\tilde \rho/\rho_{\updownarrow}) J_{s x}^0$. The corresponding minimal power dissipation density is in both cases $P_{min} = \tilde \rho \left ( J_{i x}^0 \right )^2$.

The simple but fundamental result $J_{y \updownarrow} = 0$ has been disregarded in the spin-Hall literature \cite{Dyakonov1,Hirsch,Zhang, Zutic,Maekawa,Dyakonov2,Hoffmann,Zhang14,Sinova,Saslow,Sonin,Nagaosa} because it is a property of the sole stationary states and does not depend on the details of the microscopic mechanisms. It is a direct consequence of the fact that the spin-orbit forces responsible for the deviation along the direction perpendicular to the current do not work (as it is well-known for the Lorentz force). However, as will be shown below, the result Eq.(\ref{Zero}) is valid only if charge accumulation is possible at the edges. 

The principal characteristic of the spin-Hall effect is the spin-accumulation due to spin-orbit coupling. From Eq.(\ref{Result}) (\textit{i.e.} the case $i=c$) we have, when $\sigma_{\uparrow}=\sigma_{\downarrow}$ :
\begin{equation}
\Delta \mu_{\updownarrow}(y) = -2\frac{\sigma_{so}}{\sigma}y \, \tilde \rho J_{c x}^{o}+ \Delta \mu^{ch},
\label{SpinAcc}
\end{equation}
where the constant $\Delta \mu^{ch}$ accounts for the other possible contributions to the spin-accumulation in the bulk (e.g. due to spin-flip scattering, Mott relaxation, etc). The spin-Hall angles $\theta_{SH \updownarrow}$ are defined by the relations $\tan \theta_{SH \updownarrow} = - \partial_y \mu_{\updownarrow}/\partial_x  \mu_{\updownarrow}= \mp \sigma_{so}/\sigma $  (see Figure1). This bulk spin-accumulation is well-known and has been observed in various systems \cite{Dyakonov2}. 

We discuss now a different device, in which the Hall bar is deformed according to a conformal transformation $(x,y) \rightarrow e^x e^{iy}= r e^{i \theta}$, where $r, \theta$ are the polar coordinates. We then obtain a Corbino sector (Fig.1b) or the Corbino disk (Fig.2). In the case of the Corbino angular sector of angle $\alpha$, with inner radius $r_1$ and outer radius $r_2$ shown in Fig.1b, charge accumulation is allowed. The function to minimize reads now:
 \begin{eqnarray}
 && \mathcal F^{Cor}[\vec \nabla \mu_{\updownarrow}, n_{\updownarrow},\lambda, \beta] = \int_{r_1}^{r_2} \int_0^{\alpha} [ \nonumber \\
 &&  \sigma_{\uparrow} \left(\partial_r \mu_{\uparrow} \right)^2 + \sigma_{\uparrow} \left( \frac{\partial_{\theta} \mu_{\uparrow}}{r}   \right)^2 +\sigma_{\downarrow} \left(\partial_r \mu_{\downarrow} \right)^2 + \sigma_{\downarrow} \left(\frac{ \partial_{\theta} \mu_{\downarrow}}{r} \right)^2 \nonumber  \\
&& - \lambda(\theta,r) \left(\Delta \left( \mu_{\uparrow}+\mu_{\downarrow} \right) -c_{\uparrow} \Delta n_{\uparrow} -c_{\downarrow} \Delta n_{\downarrow}+2q \frac{n}{\epsilon} \right) \nonumber \\
&&- \frac{\beta(\theta)}{r^2} ( -\sigma_{\uparrow} r \partial_r \mu_{\uparrow} -\sigma_{so} \partial_{\theta} \mu_{\uparrow} \nonumber \\
&& + \epsilon_i \left( -\sigma_{\downarrow} r \partial_r \mu_{\downarrow} +\sigma_{so} \partial_{\theta} \mu_{\downarrow}\right)-\frac{I_{r}^{i}}{\alpha h} ) \, \, ]  \, r dr d\theta,
\label{F_Cor}
\end{eqnarray}
where $i=\{c,s\}$ and $I_{i r}=I_{r \uparrow} +\epsilon_i I_{r \downarrow}$. The polar gradient takes the form $\vec \nabla = (\partial_r, \partial_{\theta}/r )$ and the Laplacian is $\Delta = \frac{1}{r}\frac{d}{dr}\left(r \frac{d}{dr}\right)+\frac{1}{r^2}\frac{d^2}{d\theta^2}$. $I_{i r} =  \alpha r h J_{i r}$ is the charge or spin current injected radially (see Fig.1b), $h$ is the thickness of the layer, and $J_{r \updownarrow}=J_{\updownarrow}.\vec e_r$ is the radial current density. The radial current is such that $r J_{r \updownarrow} = - r \sigma_{\updownarrow} \partial_r \mu_{\updownarrow} \mp \sigma_{so} \partial_{\theta} \mu_{\updownarrow}$. Proceeding as in the previous case, we obtain the result for $i=c$:
\begin{equation}
\mu_{\updownarrow}^{\alpha}(r,\theta) =-\left (\ln\left( \frac{r}{r_1} \right) \pm \frac{\sigma_{so}}{\sigma_{\updownarrow}} \theta \right) \, \tilde \rho \frac{I_{c r}}{\alpha h} - \frac{C_{\updownarrow}}{2}
\label{ResultCorSect}
\end{equation}
that leads to $J_{\theta \updownarrow} = 0$. The result is the same as the one obtained for the Hall-bar.

However, if we consider a perfect Corbino disk with $\alpha = 2 \pi$, it is equivalent to a Hall bar in which the two edges are in perfect electric contact so that charge accumulation is not allowed \cite{Rque}. The variable $n$ is fixed to zero $n_{\updownarrow} = 0$ everywhere, and the Poisson constraint reduces to the harmonic equation $\nabla^2 \mu_{\updownarrow} = 0$, whose solution is:
\begin{equation}
\mu_{\updownarrow}^{Cor}(r,\theta) = -  \frac{I_{r \updownarrow}}{2 \pi \sigma_{\updownarrow} h}\ln \left( \frac{r}{r_1} \right).
\label{ResultCor}
\end{equation}
In that case, both longitudinal and transversal stationary currents are present for the two spin-channels : $J_{r \updownarrow} = I_{r \updownarrow}/(2 \pi h r)$ and $J_{\theta \updownarrow} = \mp \sigma_{so}/\sigma_{\updownarrow} J_{r \updownarrow} $. The current lines form two opposite spirals for the two channels $\uparrow$ and $\downarrow$ (see Fig.2). The currents $\vec J_{\updownarrow}$ form the same spin-Hall angles $\theta_{SH \updownarrow}$, such that $\tan\left( \theta_{SH \updownarrow} \right)= \mp \sigma_{so}/\sigma$, with $- \vec \nabla \mu_{\updownarrow}$, as for the previous cases, but the system dissipates more than $P_{min}$. 
 \begin{figure} [h!]
   \begin{center} 
   \begin{tabular}{c}
   \includegraphics[height=9cm]{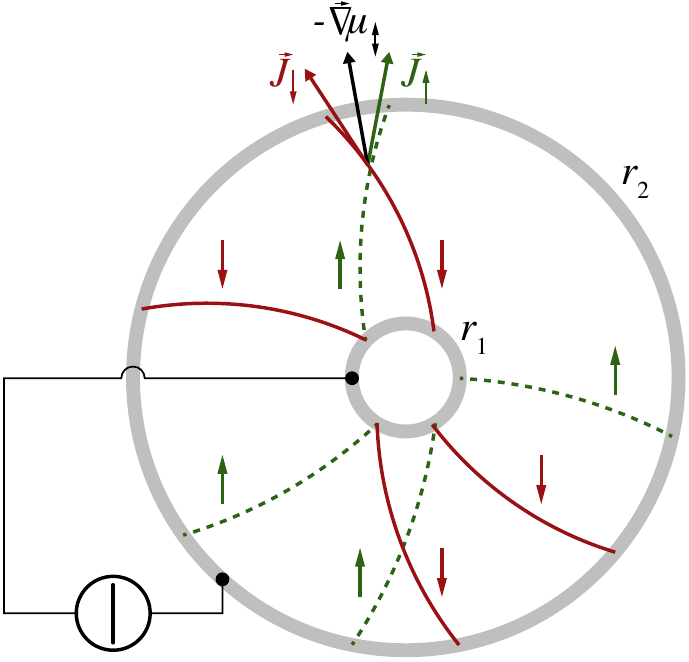}
   \end{tabular}
   \end{center}
   \caption[Fig2]
{ \label{fig:Fig2} : Schematic view of the Corbino disk with the current lines corresponding to spin $\uparrow$ and spin $\downarrow$, and $-\vec \nabla \mu_{\updownarrow}$. }
   \end{figure}
 From the experimental point of view, the most direct way to evidence the current $\vec J_{\theta \updownarrow}$ is to compare the resistances in the case of a perfect Corbino disk $R^{Cor}$ and the resistance $R^{\alpha}$ of the same scratched Corbino disk (i.e. a Corbino sector of angle $\alpha \approx 2 \pi$) . The ratio of the measured resistances is equal to the ratio of the powers \cite{SuppMat}:
  \begin{equation}
 \frac{R^{Cor}}{R^{\alpha}} = \frac{P^{Cor}}{P^{\alpha}} = 1+ \left ( \frac{\sigma_{so}}{\sigma} \right)^2
 \label{Expe}
 \end{equation}
 If we consider spin-flip relaxation, the results presented above about a bulk material are not changed in a device larger than the spin-diffusion length. Indeed, in case of spin-flip scattering it is necessary to add the spin-flip dissipation in the expression of the power $P_J^{sf} = P_J + \dot \psi \Delta \mu^{ch}$, where $\Delta \mu^{ch} = \mu_{\uparrow}^{ch} - \mu_{\downarrow}^{ch}$, and $\dot \psi$ is the spin-flip current in the spin configuration space (we use here the notation of the chemical reaction that transforms electric carriers of spin $\uparrow$ into electric carriers of spin $\downarrow$ \cite{MagDiff,Entropy}). The corresponding supplementary transport equation reads $\Dot \psi = L \Delta \mu^{ch}$, where $L$ is a constant transport coefficient. The supplementary variable $\dot \psi$ (or $\Delta \mu^{ch}$) has to be considered in the functions $\mathcal F$ and $\mathcal F^{Cor}$(Eq.(\ref{F_Min}) and Eq.(\ref{F_Cor})). However, the minimization leads to the condition $\dot \psi = 0$. In other terms, the two spin-channels are at equilibrium with respect to the spin-flip relaxation in the bulk. As a consequence, the results obtained above under the assumption of zero spin-flip scattering can be generalized to DP model including spin-flip scattering.
 
In conclusion, we have shown that the spin to charge current conversion can be performed in the case of the Corbino geometry as far as charge accumulation is not allowed. The conversion equation $\vec J_c= (- \sigma/\sigma_{so}) \vec J_s \times \vec p$ is then verified \cite{SuppMat}. In contrast, the spin to charge current conversion at stationary state cannot be performed if charge accumulation occurs, in particular for the usual Hall-bar geometry or for the scratched Corbino disk. In those configurations, the above conversion relation between $\vec J_c$  and $\vec J_s$  is not verified. These predictions could be compared experimentally by measuring the ratio of the resistances $1+ (\sigma_{so}/\sigma)^2$ of the same scratched and non-scratched Corbino disk.
\section{acknowledgment}
  JEW thanks A. Fert and M. Dyakonov for fruitful discussions.
  R.B. was sponsored by the Chair X-ESPCI-Saint Gobain.


\begin{thebibliography}{47}

 
\bibitem{Dyakonov1} M. I. Dyakonov, and  V. I. Perel, JETP Lett. {\bf 13}, 467"1¤7470 (1971)
and M. I. Dyakonov, and  V. I. Perel, Phys. Lett. {\bf A 35}, 459"1¤7460 (1971).
\bibitem{Hirsch} J. E. Hirsch, Phys. Rev. Lett. {\bf 83}, 1834"1¤71837 (1999). 
\bibitem{Zhang} S. Zhang, Phys. Rev. Lett.
{\bf 85}, 393"1¤7396 (2000).
\bibitem{Zutic} W.-K. Tse, J. Fabian I. $\check{Z}$uti\'c, and S. Das Sarma, Phys. Rev. B {\bf 72}, 241303(R) (2005).
\bibitem{Maekawa} S. Takahashi and S. Maekawa, Sci. Technol. Adv. Mater. {\bf 9} 014105 (2008)
\bibitem{Dyakonov2} M. I. Dyakonov spin Physics in Semiconductors, Springer Series in Solid-States Sciences 2008.
\bibitem{Hoffmann} A. Hoffmann, IEEE Trans. Mag. {\bf 49} (2013) 5172.
\bibitem{Zhang14} S. S.-L. Zhang, K. Chen, and S. Zhang, Europhys. Lett. {\bf 106}, 67007 (2014).
\bibitem{Sinova} J. Sinova, S. O. Valenzuela, J. Wunderlich, C. H. Back, T. Jungwirth, Rev. Mod. Phys. {\bf 87}, 1213 (2015).
\bibitem{Saslow} W. M. Saslow, Phys. Rev. B {\bf 91}, 014401 (2015).
\bibitem{Patent} K Uchida, K Harii, Y Kajiwara, E Saitoh , US Patent 8203191 B2 (2012).
\bibitem{Nagaosa} N. Nagaosa, J. Sinova, S. Onoda, A. H. MacDonald, N. P. Ong,  Rev. Mod. Phys. {\bf 82}, 1539 (2010). 
\bibitem{Jedema} F. J. Jedema, A. T. Filip and B. J. van Wees, Nature {\bf 410}, 345 (2001)
\bibitem{Georges} J.-M. Georges, A. Fert, G. Faini, Phys. Rev. B {\bf 67}, 012410 (2003).
 \bibitem{Otani} Y. Omori, F. Auvray, T. Wakamura, Y. Niimi, A. Fert, and Y. Otani, Appl. Phys. Lett. {\bf 104}, 242415 (2014).
\bibitem{Sonin} E. B. Sonin Adv. Phys. {\bf 59}, 181-255 (2010)
\bibitem{AThT} J.-E. Wegrowe, H.-J. Drouhin, D. Lacour, Phys. Rev. B {\bf 89}, 094409 (2014)
\bibitem{Onsager1} L. Onsager, Phys. Rev. {\bf 37} 405 (1931).
\bibitem{Onsager2} L. Onsager, Phys. Rev. {\bf 38} 2265 (1931).
\bibitem{Rubi} D. Reguera, J. M. G. Vilar, and J. M. Rub\`i, J. Phys. Chem. B 109 (2005).
\bibitem{MagDiff} J.-E. Wegrowe, S. N. Santos, M.-C. Ciornei, H.-J. Drouhin, J. M. Rub\'i, Phys. Rev. B {\bf 77} , 174408 (2008).
\bibitem{Entropy} J.-E. Wegrowe, H.-J. Drouhin, Entropy, 13, 316 (2011). 
\bibitem{SPIE14} J.-E. Wegrowe et al. Spintronics VII, Proceeding of the SPIE {\bf 9167}, 91671O-1 (2014).

\bibitem{Onsager3} L. Onsager and S. Machlup, Phy. Rev. {\bf 91} 1505 (1953).
\bibitem{Jaynes} E. T. Jaynes, Ann. Rev. Phys. Chem. {\bf 31} 579 - 601 (1980).

\bibitem{Rque} Charge accumulation is still allowed at the surface of a conducing wire of size much larger than $\lambda_D$ that connects the two edges of the Hall bar. 


\bibitem{SuppMat} Supplementary materials. 





 
\end{thebibliography}
 \end{document}